\documentclass[paper,prd,reprint,preprintnumbers,nofootinbib,notitlepage,showpacs,showkeys]{revtex4-2}
\usepackage{latexsym,graphicx,amssymb,amsmath,mathrsfs} 
\usepackage{slashed,mathbbol}
\usepackage[utf8]{inputenc}

\DeclareSymbolFont{bbold}{U}{bbold}{m}{n}
\DeclareSymbolFontAlphabet{\mathbbold}{bbold}

\newcommand{\be}{\begin{equation}}      
\newcommand{\ee}{\end{equation}}      
\newcommand{\bea}{\begin{eqnarray}}      
\newcommand{\eea}{\end{eqnarray}}    
\newcommand{\Tr}{\,\textrm{Tr}\,}
\newcommand{\cl}{\,\textrm{cl}\,}

\newcommand{\ns}{\,\textrm{ns}\,} 
\newcommand{\s}{\,\textrm{s}\,} 
\newcommand{\MeV}{\,\textrm{MeV}\,} 
\newcommand{\GeV}{\,\textrm{GeV}\,} 
\newcommand{\eff}{\,\textrm{eff}\,}

\newcommand{\disc}{\,\textrm{disc}\,} 
\newcommand{\un}{\mbox{$\mathbf{1}$   
\hspace{-0.91em}

\raisebox{0.05em}[0pt]{$\shortmid$}   
\hspace{-0.895em} \raisebox{0.235em}[0pt]{$\shortmid$}}}

\makeatletter
\renewcommand\appendix{\par
\setcounter{section}{0}%   
\setcounter{subsection}{0}% 
\gdef\thesection{\appendixname\space\@Alph\c@section}}

\long\def\unmarkedfootnote#1{{\long\def\@makefntext##1{##1}\footnotetext{#1}}}
\makeatother

\begin{document} 

\title{Thermal behavior of effective $U_A(1)$ anomaly couplings \\in reflection of higher topological sectors} 
\author{G. Fej\H{o}s}
\email{gergely.fejos@ttk.elte.hu}
\affiliation{Institute of Physics and Astronomy, E\"otv\"os University, 1117 Budapest, Hungary}
\affiliation{RIKEN iTHEMS, Wako, Saitama 351-0198, Japan}
\author{A. Patk\'os}
\email{patkos@galaxy.elte.hu}
\affiliation{Institute of Physics and Astronomy, E\"otv\"os University, 1117 Budapest, Hungary}

\begin{abstract}
{Thermal behavior of effective, chiral condensate-dependent $U_A(1)$ anomaly couplings is investigated using the functional renormalization group approach in the $N_f = 3$ flavor meson model. We derive flow equations for anomaly couplings that arise from instantons of higher topological charge, dependent also on the chiral condensate. These flow equations are solved numerically for the $|Q|=1,2$ topological sectors at finite temperature. Assuming that the anomaly couplings at the ultraviolet scale may also exhibit explicit temperature dependence, we calculate the thermal behavior of the effective potential. In accordance with our earlier study, [G. Fejos and A. Patkos, Phys. Rev. D{\bf 105}, 096007 (2022)], we find that for increasing temperatures, the anomalous breaking of chiral symmetry tends to strengthen toward the pseudocritical temperature ($T_C$) of chiral symmetry breaking. It is revealed that below $T_C$, around $\sim$10\% of the $U_A(1)$ breaking arises from the $|Q|=2$ topological sector. Correspondingly, a detailed analysis on the thermal behavior of the mass spectrum is also presented. }
\end{abstract}

\maketitle

\section{Introduction}

The axial $U_A(1)$ subgroup of the approximate $U_V(N_f)\times U_A(N_f)$ chiral symmetry of quantum chromodynamics (QCD) with $N_f$ quark flavors is known to be broken anomalously in the ground state of the system \cite{schaefer96,schaefer98}. Microscopic origin of this anomaly, at least at sufficiently high temperatures, can be well described in the dilute instanton gas picture. However, at lower temperatures, toward the pseudocritical point ($T_c$), the eigenvalue spectrum of the Dirac operator is rather different to its high $T$ counterpart, revealing that the instanton gas approximation definitely breaks down \cite{ding21a,vig21}. The correct picture in such regimes is assumed to be more of an instanton liquid \cite{shuryak82}, where the instanton density and radius play a central role. Though the understanding of the underlying mechanism of the $U_A(1)$ breaking has significantly improved over time, the actual fate of the anomaly along the complete temperature axis still remains an open question.

Since the isotriplet pseudoscalar meson ($\pi$) and the isotriplet scalar meson ($a_0$) are related through the axial $U_A(1)$ transformation, it is natural to quantify the degree of the anomaly breaking by either the $m_{a_0}-m_{\pi}$ mass or $\chi_{\pi}-\chi_{a_0}$ susceptibility differences. In a chirally symmetric background, where the $\pi$ and $\sigma$ excitations degenerate, one notes that the former difference is equivalent to the disconnected part of the total chiral susceptibility, $\chi_{\pi}-\chi_{a_0} = \chi_{\pi}-\chi_{\sigma}+\chi_{\disc} \rightarrow \chi_{\disc}$ \cite{kaczmarek20}, leaving the quantification of the anomaly to determine $\chi_{\disc}$. 

Lattice QCD studies have not yet reached a firm conclusion on the presence of the anomaly at the pseudocritical point.  Refs. \cite{bazazov12,buchoff14,bhattacharya14}, with the use of domain-wall fermions, concluded that the anomaly is present even beyond the pseudocritical temperature. Ref. \cite{brandt16}, using Wilson fermions, shows that $U_A(1)$ symmetry is effectively restored at $T_c$ in the chiral limit (for $N_f=2$).  Eigenvalue spectrum analyses argue that the anomaly is present even beyond the critical temperature \cite{dick15}, and using highly improved staggered quark action also shows that $U_A(1)$ is broken at $1.6 T_c$ \cite{ding21}.  The same finding is presented in Ref. \cite{kaczmarek21} just above $T_c$, using analyses of eigenvalue densities. Ensembles generated by two-flavor (Möbius) domain wall sea quarks with the eigenvalue analysis show,  however, that the obtained results are consistent with the $U_A(1)$ symmetry being restored at $T_c$, at least in the chiral limit \cite{tomiya16,aoki21}. Predictions of the density of state method in the $SU(3)$ pure gauge theory show that even the $|Q|=2$ topological sector has a non-negligible contribution to the topological susceptibility even at $1.5T_c$ \cite{borsanyi21}. For a comprehensive review on recent developments along these directions, the reader is referred to Ref. \cite{lahiri21}.

On top of lattice QCD simulations, there are several other directions to tackle the problem of the thermal evolution of the anomaly.  See, e.g.,  studies using the nonlinear sigma model, chiral perturbation theory, Polyakov quark meson model \cite{kovacs16,rennecke16,gomeznicola16,rai20,li20,gomeznicola21,tiwari23}, NJL models \cite{ishii16,ishii17}, the Witten-Di Vechia-Veneziano model \cite{bottaro20}. There have been attempts using the Dyson-Schwinger approach \cite{horvatic19,horvatic20}, exploiting Ward identities \cite{gomeznicola19}. One may also be interested in investigating the issue in two-color QCD with $N_f=2$ flavors \cite{kawaguchi23}.

The order of the chiral transition for zero quark masses may also contain information on the $U_A(1)$ restoration at the critical temperature. Recently, there have been indications that the chiral phase transition is of second order for both $N_f=2,3$ in the chiral limit \cite{cuteri21,dini22,bernhardt23} in contradiction with earlier studies \cite{chandrasekharan07}, especially those using the perturbative renormalization group in the $\epsilon$ expansion \cite{pisarski84}. The latter shows that for $N_f=3$, irrespectively of the fate of the $U_A(1)$ breaking at the critical point, the transition can only be of first order. Recently, using the functional renormalization group (FRG) technique, however, it was shown that the transition can appear to be second order after all, but only if the anomaly is very weak at $T_c$ \cite{fejos22}.

In our earlier studies \cite{fejos16,fejos21} we argued that in the $N_f=3$ low energy effective meson model, fluctuations tend to make the anomaly stronger with respect to the temperature. The newly found mechanism is related to the resummation of an infinite number of anomaly breaking operators in the functional renormalization group (FRG) formalism, which effectively made the standard Kobayashi-Maskawa-'t Hooft (KMT) coupling chiral condensate dependent. Through such dependence one observed a strengthening of the effective anomaly coupling as the condensate evaporated, showing that mesonic fluctuations are working against instanton contributions, which, at least at asymptotically large $T$ definitely lead to the restoration of $U_A(1)$.

The standard KMT coupling and the corresponding determinant term arise from instanton contributions of the underlying theory that carry $Q=\pm 1$ topological charge. In \cite{pisarski20} it is explicitly shown for $N_f=2$ that instantons with higher winding numbers lead to higher powers of the determinant term, which are typically considered nonrenormalizable, and thus entirely dropped in the effective description. Even though one might expect that the value of these couplings in the ultraviolet (even being zero) does not effect significantly the physics of the infrared, the corresponding terms do get generated at low energies, and in principle should not be neglected in the effective action. The main goal of this study is to derive scale evolution equations for those anomalous terms that arise from instantons with any winding number, and on top of that, realize a resummation in terms of chirally invariant operators to make the former condensate dependent. Our aim is also to quantify the importance of the higher topological sectors in the effective model setting. 

We wish to emphasize that a consistent treatment of the $U_A(1)$ breaking, at least considering the physical point, can only be dealt with via the $N_f=3$ scenario.  In case of $N_f=2$, the strange sector is completely decoupled as the $s$-quark mass is formally taken to be infinity. In such a case, the determinant in effect becomes a mass term, and the corresponding anomaly parameter also needs to be set to infinity in order for one of the $O(4)$ multiplets can be dropped, as required by, e.g. the $m_K < m_{\eta'}$ mass relation in the physical point.  (Keep in mind that because of suppressing the strange content,  by consistency $m_K \rightarrow \infty$, thus $m_{\eta'} \rightarrow \infty$.) Even if one wishes to keep both multiplets \cite{pisarski20,grahl13}, one faces the fact that beyond $T_c$, by the effective evaporation of the (nonstrange) chiral condensate, keeping track of anomalous contributions in the effective potential is no longer possible. The $N_f=3$ model, however, does, through the very presence of the strange condensate, allow to investigate the effective KMT coupling even beyond $T_c$, which proves to be crucial in understanding characteristic features of the high temperature meson spectrum.

The paper is organized as follows. In Sec. II, we discuss our model setup and introduce the corresponding ansatz for the effective potential, with particular emphasis on the resummation that leads to condensate dependent effective couplings. In Sec. III, we discuss how to obtain the scale evolution of the complete effective potential via the use of various background fields for the construction of the FRG flow equations (presented explicitly in the Appendix). Section IV contains the numerical results, where we used a model parameter set that corresponds to the physical point.  The reader also finds a detailed analysis on the temperature dependence of the mesonic spectrum in light of the anomaly evolution. Section V is devoted for summary.

\section{Model setup}

Our investigations involve a mesonic dynamical variable $M=(s_i+i\pi_i)T_i$, where $T_i$ $(i=0,...,8)$ are the $U(3)$ generators, and $s_i$ ($\pi_i$) refer to the scalar (pseudoscalar) fields. A chiral transformation is represented as $M \rightarrow LMR^\dagger$, where $L$ ($R$) are left (right) handed transformations. The effective potential, $V$, of our model can only depend on chirally invariant combinations of $M$. For $N_f=3$, a possible set of independent operators are the usual
\begin{subequations}
\label{Eq:inv}
\bea
\label{Eq:rho}
\rho&=& \Tr (M^\dagger M), \\
\label{Eq:tau}
\tau &=& \Tr (M^\dagger M - \Tr(M^\dagger M)/3)^2, \\
\label{Eq:rho3}
\rho_3 &=& \Tr(M^\dagger M-\Tr(M^\dagger M)/3)^3
\eea
\end{subequations}
combinations. Higher order invariants can be expressed in terms of (\ref{Eq:inv}). The $U_A(1)$ breaking can be included via the
\bea
\label{Eq:Delta}
\Delta = \det M^\dagger + \det M
\eea
Kobayashi-Maskawa-'t Hooft (KMT) term, where it is important to mention that the plus sign in the rhs of (\ref{Eq:Delta}) is due to parity reasons (a minus sign would lead to a parity odd combination), and also that $\tilde{\Delta}\equiv (\det M^\dagger - \det M)^2$ is not independent from (\ref{Eq:Delta}) and (\ref{Eq:inv}). That is, the potential, both classical or quantum, can only depend on $\rho,\tau,\rho_3,\Delta$. Note that in the broken phase chiral symmetry shows the $U(3)\times U(3) \rightarrow U(3)$ breaking pattern, which is realized by the condensate $\langle M \rangle \sim \un$, i.e. it is proportional to the unit matrix. In such backgrounds $\tau = 0 = \rho_3$, i.e., even if one includes explicitly symmetry breaking terms (arising from nonzero quark masses) that somewhat shift the vacuum expectation value of $M$, it is expected that neither $\tau$ nor $\rho_3$ carry significant contributions in $V$. Therefore, it is expected to make sense to perform the following chiral invariant expansion \cite{fejos14}:
\bea
\label{Eq:V}
V(\rho,\tau,\rho_3,\Delta) = U(\rho,\Delta) + \sum_{\{\alpha\}} V_{\alpha_1,\alpha_2} (\rho,\Delta)\tau^{\alpha_1} \rho_3^{\alpha_2},
\eea
where $\alpha_1, \alpha_2 \in \mathbb{Z}$. Note that in the classical version of the model, based on perturbative renormalizability, $\rho_3$ is dropped, and an expansion in terms of $\rho$ and $\Delta$ is also performed:
\bea
\label{Eq:Vcl}
V_{\cl} = m^2 \rho + \lambda_1 \rho^2 + a \Delta +...+ \lambda_2 \tau + ...,
\eea
which is the usual potential of the three flavor linear sigma model with real parameters $m^2, \lambda_1, \lambda_2, a$. In the original variable $M$, (\ref{Eq:Vcl}) yields
\bea
\label{Eq:VclM}
V_{\cl} &=& m^2 \Tr(M^\dagger M) + \tilde{\lambda}_1[\Tr(M^\dagger M)]^2 \nonumber\\
&+& \lambda_2 \Tr(M^\dagger M M^\dagger M) + a (\det M^\dagger + \det M),
\eea
with $\tilde{\lambda}_1=\lambda_1-\lambda_2/3$. Note that at high momentum scales the form of (\ref{Eq:VclM}) is certainly correct\footnote{We note that recently it has been conjectured that in the chiral limit the $Q>1$ anomaly couplings may dominate the $Q=1$ one \cite{pisarski24}. For our case it would mean that $a$ is negligible and one would have to start the RG flow with the inclusion of the $\sim \Delta^2$ operator at the UV scale.}, but for low scales it is very restrictive, and thus if one includes quantum and thermal fluctuations, the more general form, (\ref{Eq:V}) should be considered, at least via some approximation. From here onward, we entirely drop the $\rho_3$ dependence due to the reason described above, but we do keep $\tau$, as it corresponds to a renormalizable operator that should be present at the UV scale. Its coefficient is, in principle, $\rho$ and $\Delta$ dependent, but we consider only the former:
\bea
\label{Eq:Vansatz}
V = U(\rho,\Delta) + C(\rho) \tau,
\eea
which will be the main starting point of our investigations. Even though (\ref{Eq:Vansatz}) looks very simple, it is only due to our compact notations. Note that, the infinite resummation (\ref{Eq:Vansatz}) realizes is threefold:
\bea
\label{Eq:Vsum}
V &=& \sum_{\beta} U^{(\beta)}(\rho) \Delta^\beta + C(\rho)\tau \nonumber\\
&\equiv& \sum_{\alpha} \sum_{\beta} U^{(\alpha,\beta)} \rho^\alpha \Delta^{\beta}+ \sum_{\alpha} C^{(\alpha)} \rho^\alpha \tau.
\eea
Here $U^{(\alpha,\beta)}$ and $C^{(\alpha)}$ can be thought of as usual coupling constants. This demonstrates that the simple form of (\ref{Eq:Vansatz}) contains an infinite number of couplings. Our goal is to determine these couplings through obtaining the functions $U(\rho,\Delta)$ and $C(\rho)$.

A comment on the anomalous terms is now in order. It was shown by Pisarski and Rennecke that the $\sim \!\Delta^\alpha$ terms in the potential arise from multi instanton contributions of the underlying theory, with $|Q|=\alpha$ topological charges. Emergence of these interactions for $N_f=2$ can be found in \cite{pisarski20}. In what follows, we investigate the importance of the interactions generated by multi instanton contributions, in terms of an effective mesonic description at finite temperature.  In the next section, using the functional renormalization group (FRG) formalism, we derive and then solve scale evolution equations for $U(\rho,\Delta)$ and $C(\rho)$.

\section{RG flows}

In the core of the FRG formalism lies the scale dependent effective action ($\Gamma$), which, throughout this paper will be considered in the local potential approximation (LPA). That is, $\Gamma$ only contains a standard kinetic term besides the effective potential, $V$. Denoting the scale variable by $k$, the latter obeys the Wetterich equation \cite{wetterich93,morris94}, which, in $d$ spatial dimensions, using Litim's regularization \cite{litim01} takes the form of
\bea
\label{Eq:wet}
\partial_k V_k &=& T\Omega_d \frac{k^d}{2d}\sum_{n} \tilde{\partial}_k \log\det (\omega_n^2 + k^2 + V_k''),
\eea
where $T$ is the temperature, $\Omega_d = \int d\Omega_d/(2\pi)^d$ comes from the angular integral, $\omega_n=2\pi nT$ are bosonic Matsubara frequencies, $V_k''$ is the mass matrix, meaning the second derivative of $V_k$ with respect to the dynamical variables (in our case, i.e., for $N_f=3$ flavors its size is $18\times 18$), and $\tilde{\partial}_k$ is a differential operator acting by definition only on the explicit $k$-dependence. Note that, the physical meaning of $V_k$ is an effective potential, where only fluctuations with momenta $q>k$ are taken into account. Practically $(\ref{Eq:wet})$ is solved by starting from a typical hadronization scale $\Lambda$ \cite{mitter15}, where $V_{k\rightarrow \Lambda}$ is considered to be the fluctuationless classical potential, and we integrate down toward $k\rightarrow 0$. We also note that (\ref{Eq:wet}) is derived with an infrared regulator that only cuts off fluctuations in the spatial directions, thus the Matsubara sum runs over all possible frequencies.

Since we are working with the ansatz of (\ref{Eq:Vansatz}), we need to transform the right-hand side (rhs) of (\ref{Eq:wet}) such that it becomes compatible with (\ref{Eq:Vansatz}). We follow the same procedure developed in \cite{fejos16,fejos21}. The most important observation is that in order to derive flow equations for $U_k(\rho,\Delta)$ and $C_k(\rho)$, one is allowed to work in any background field of $\langle M \rangle$ that allows for a unique reconstruction of the dependence on the invariants, not necessarily the one, which will be realized physically in the ground state.

We start by calculating the flow of $U_k(\rho,\Delta)$. Since for any $\langle M \rangle \sim \un$, $\tau = 0$, in such a background the left-hand side (lhs) of (\ref{Eq:wet}) equals the flow of $U_k(\rho,\Delta)$. (Note that, the derivatives of $\tau$ are, in principle, nonzero in this background.) We, therefore, choose $\langle M \rangle = (s_0+i\pi_0)T_0$, $T_0=1/\sqrt6 \un$. In this case, the invariants become $\rho=(s_0^2+\pi_0^2)/2$ and $\Delta=s_0(s_0^2-3\pi_0^2)/(3\sqrt6)$. Note that, this background respects the $SU(3)\times SU(3)$ subgroup of chiral symmetry, therefore, the mass matrix $V_k''$ has 8+8 degenerate eigenmodes in the ``planes'' $(s_i,\pi_i) (i=1,...,8)$, and a doublet in the $(s_0,\pi_0)$ sector. It is important to note that throughout the calculation each matrix element depends on the background components of $s_0$, $\pi_0$, but in the 8+8 degenerate eigenmodes they naturally combine into the $\rho$ and $\Delta$ invariants, see their respective expressions above. This is not true for the doublet, but after calculating the corresponding $2\times 2$ determinant, its contribution to the flow equation happens to depend again only on the invariant combinations $\rho$ and $\Delta$, as it should. Summing up all terms yields the flow for $U_k(\rho,\Delta)$, see (\ref{Eq:flowU}) in the Appendix.

For determining the scale evolution of the coefficient function of the $\tau$ invariant, that is $C_k(\rho)$, the purely imaginary background $\langle M \rangle = i(\pi_0T_0+\pi_8T_8)$ is particularly convenient. In this background the cubic invariant $\Delta$ vanishes, but $\rho= \frac{1}{2}(\pi_0^2+\pi_8^2)$ and $\tau=\frac{\pi_8^2}{3}\big(\pi_0-\frac{1}{2\sqrt{2}}\pi_8\big)^2$. That is, the lhs of (\ref{Eq:wet}) now becomes $\partial_k U_k(\rho,0)+\partial_k C_k(\rho) \tau$, and since we already obtained $\partial_k U_k$, by calculating the rhs of (\ref{Eq:wet}) and subtracting the $\partial_k U_k$ part, we obtain the flow of $C_k(\rho)$.  In accordance with \cite{fejos21}, the mass matrix breaks up into three degenerate $\{\sigma_i,\pi_i\}$ doublets ($i=1,2,3$), plus another four degenerate doublets ($i=4,5,6,7$), with a fully coupled quartet in the subspace \{$s_0,s_8,\pi_0,\pi_8$\}. One way to proceed with the calculations is that one inverts the $\rho$ and $\tau$ equations and expresses $\pi_0, \pi_8$ in terms of the latter. Then all terms in the rhs of the flow equation can be expressed with the help of the invariants $\rho$ and $\tau$, but at first sight they turn out to be non analytic in the latter, i.e. they contain terms $\sim \!\sqrt{\tau}$. When expanding the contribution of each sector in terms of $\tau$, the piece that comes from the \{$s_0,s_8,\pi_0,\pi_8$\}-sector completes the two sets of degenerate doublets into an analytic function of both $\rho$ and $\tau$, meaning that all the peculiar $\sim\!\sqrt{\tau}$ terms vanish. After a long and tedious calculation, the first nontrivial order in $\tau$ provides the flow for $C_k(\rho)$, see (\ref{Eq:flowC}) in the Appendix.

\section{Results}

\subsection{Numerics}

\begin{table}[t]
\centering
\vspace{0.2cm}
  \begin{tabular}{ c | c }
    $m^2$ & $-0.95 \GeV^2$  \\ \hline
    $\lambda_1$ & 22.7 \\ \hline
    $\lambda_2$ & 130 \\ \hline
    $a$ & $-2.4 \GeV$ \\ \hline  
    $h_0$ & $ (285 \MeV)^3$ \\ \hline  
     $h_8$ & $(-310 \MeV)^3$ \\
  \end{tabular}
  \caption{Initial conditions at $k=\Lambda\equiv 1 \GeV$.}
\end{table}

We do not intend to solve the flow equation for $U_k(\rho,\Delta)$ in its full generality, mainly due to the high numerical cost. In what follows we are interested in the scale evolution of the $|Q|=1,2$ instanton contributions. By introducing the notations $U(\rho):= U^{(0)}(\rho)$, $A(\rho):= U^{(1)}(\rho)$, $B(\rho):= U^{(2)}(\rho)$, using the decomposition of (\ref{Eq:Vsum}), we consider the effective potential as
\bea
\label{Eq:Vfinal}
V_k = U_k(\rho) + C_k(\rho)\tau + A_k(\rho) \Delta + B_k(\rho) \Delta^2.
\eea
The flow equation for $C_k(\rho)$ is given by (\ref{Eq:flowC}), while those for $U_k(\rho)$, $A_k(\rho)$ and $B_k(\rho)$ can be obtained by the zeroth, first, and second order terms in a $\Delta$ expansion of (\ref{Eq:flowU}). The choice of (\ref{Eq:Vfinal}) allows us to investigate the importance of the $|Q|=2$ interaction [i.e. $B_k(\rho)$] in light of the one with $|Q|=1$ [i.e. $A_k(\rho)$].

The coupled differential equations are solved for three spatial dimensions ($d=3$) using the grid method. All grids are set up in $\rho$ space with the spacing $\delta \rho= 50\MeV^2$. Field derivatives are calculated using the six-point formula, except those close to the boundaries, where the five- and four-point formulas are used. Initial conditions are chosen to correspond to $k=\Lambda\equiv 1 \GeV$, where, based on perturbative renormalizability, we assume that
\bea
\label{Eq:init}
U_{k=\Lambda}(\rho)&=& m^2 \rho + \lambda_1 \rho^2, \quad C_{k=\Lambda}(\rho) = \lambda_2, \nonumber\\
A_{k=\Lambda}(\rho)&=&a, \quad B_{k=\Lambda}(\rho)=0,
\eea
with some real UV parameters $m^2$, $\lambda_1$, $\lambda_2$, $a$. These initial values at $k=\Lambda$ are determined by the requirement of optimally reproducing the pseudoscalar meson spectra once all fluctuations are taken into account, see Table I. Note that, the initial condition for the $B_k(\rho)$ coefficient function is set to be zero, which is due to the (perturbative) irrelevance of the $\Delta^2$ operator at high scales. We have checked the sensitivity of the results with respect to the variation of the UV value of the latter interaction in the natural range of ${\cal O}([0.1-10]/\Lambda^2)$, and we have found no significant difference between the final results, as expected. Note that, in the physical point one needs to subtract an explicit symmetry breaking term from the potential, i.e. $h_0T_0 + h_8T_8$, where $h_0$ and $h_8$ are determined using the partially conserved axialvector current relations (PCAC), see details in \cite{fejos16,fejos21}. As discussed earlier, without explicit breaking, in the vacuum $\langle M \rangle \sim T_0 \sim \un$, but by introducing $h_0$ and $h_8$ into the potential, the minimum point of $V$ shifts, and the actual physical background becomes $\langle M \rangle = v_0T_0+v_8T_8$.

The differential equations are integrated via the fourth-order Runge-Kutta method. We typically expect a critical slowing down in the numerics when approaching $k\rightarrow 0$, therefore, we stop the flows at $k/\Lambda=0.1$. At this point we avoid numerical instabilities, while all functions are practically converged and the scale dependence is rather moderate. We also note that all Matsubara sums are performed numerically, with a cutoff of $n_{\max}=10^4$.

\subsection{Ground state and anomaly}

All results are split into two parts. On the one hand, we treat the initial $a$ parameter (which corresponds to the bare parameter in the perturbative renormalization process) as a constant, but on the other hand, we are also interested in a scenario, where it explicitly depends on the temperature.  The latter is motivated by the fact that since we are dealing with an effective theory cut off at the rather low $\Lambda = 1\GeV$ scale, the initial parameters should inherit non-negligible environment dependence when deriving them from the underlying theory of QCD.  Denoting by $T_c$ the pseudocritical temperature of the chiral transition, it is known that in the dilute instanton gas approximation, valid for $T\gg T_c$, the instanton density is exponentially suppressed, leading to recovery of the anomalously broken axial $U_A(1)$ symmetry.  When lowering the temperature, however, an instanton liquid model is more appropriate, and it is sometimes argued that the instanton density is weakly dependent on the temperature below $T_c$ \cite{shuryak94}. Therefore, as a working hypothesis, and motivated by obtaining the correct large $T$ behavior of the anomaly, we make the initial anomaly parameter of the meson model temperature dependent as
\bea
\label{Eq:aT}
\!\!\!\!\!a(T)=a(T=0)\left[1+\left(\left(\frac{T_c}{T}\right)^{5}-1\right)\Theta(T-T_c)\right].
\eea
This choice activates the high-temperature behavior of the anomaly beyond the critical temperature and maintains it constant below the critical temperature. For $T>T_c$, the $a \sim T^{-5}$ behavior originates from integrating over the instanton size after applying the semi classical approximation for the instanton density \!\!\footnote{Note that had one retained a nonzero initial value for $B_{k=\Lambda}\equiv b$, for the same reasons as described in the main text, one would have had $b(T)=b(T=0)\left[1+\left(\left(T_c/T\right)^{14}-1\right)\Theta(T-T_c)\right]$.} \cite{pisarski20}. Note that, the remaining $m^2$, $\lambda_1$, $\lambda_2$ parameters can, in principle, also carry explicit $T$-dependence, but for the sake of our main motivation of investigating the anomaly evolution, these parameters will be treated as temperature independent quantities in the current study.

\begin{figure}[t]
\includegraphics[bb = 350 80 495 590,scale=0.33]{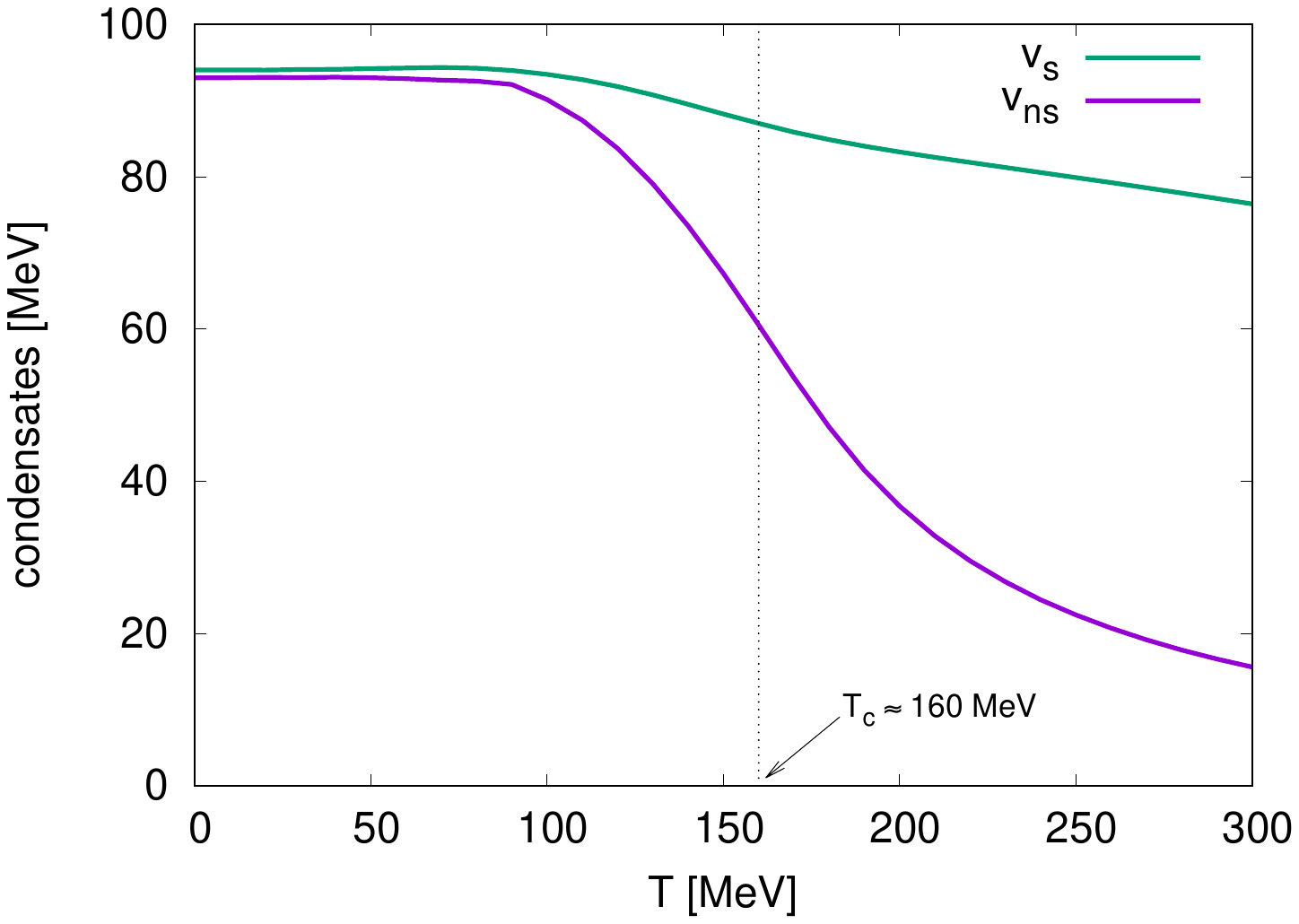}
\caption{Thermal behavior of the nonstrange and strange condensates with a temperature {\it independent} initial anomaly parameter. Prediction for the pseudocritical temperature is $T_c \approx 160 \MeV$.}
\label{Fig:cond1}
\includegraphics[bb = 350 80 495 590,scale=0.33]{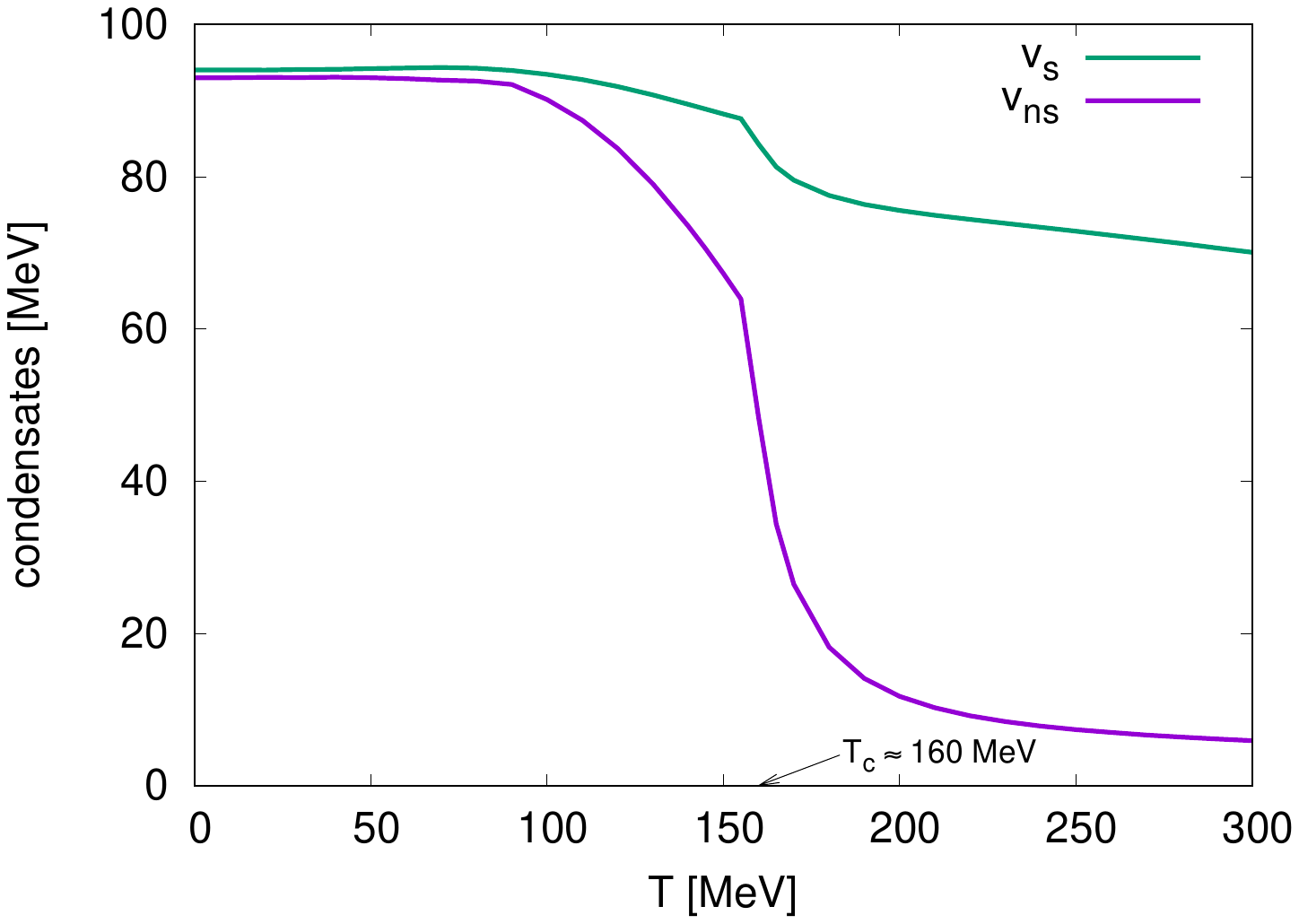}
\caption{Thermal behavior of the nonstrange and strange condensates with a temperature {\it dependent} initial anomaly parameter. Instanton contributions are activated exactly at $T=T_c$.}
\label{Fig:cond2}
\end{figure} 

In Fig. \ref{Fig:cond1} and Fig. \ref{Fig:cond2} we show the thermal evolution of the ground state (i.e. the minimum point of the effective potential $V$) in terms of the nonstrange ($v_{\ns}$) and strange ($v_{\s}$) condensates for both scenarios described above.\!\footnote{The nonstrange and strange components are defined through the usual $v_{\ns} = \sqrt{\frac23} v_0 + \sqrt{\frac13} v_8$, $v_{\s}=\sqrt{\frac13}v_0 - \sqrt{\frac23}v_8$ relations.} The prediction for the pseudocritical temperature, defined as the inflection point of the $v_{\ns}(T)$ function, is $T_c \approx 160 \MeV$, being very close to lattice simulations \cite{aoki06,bazazov12}.  The break point in case of the second scenario has no physical meaning, it is due to the fact that $a(T)$ is a nondifferentiable function at $T=T_c$. The choice we made for the activation point of the $T$-dependence is somewhat arbitrary, and furthermore, had we chosen a smoother interpolation between the two regimes, we would have obtained a curve without breaking points.  Note that while the nonstrange condensate shows a significant evaporation, the strange component loses less than $20\%$ of its $T=0$ value at $T\approx 2T_c$.

\begin{figure}[t]
\includegraphics[bb = 350 80 495 590,scale=0.33]{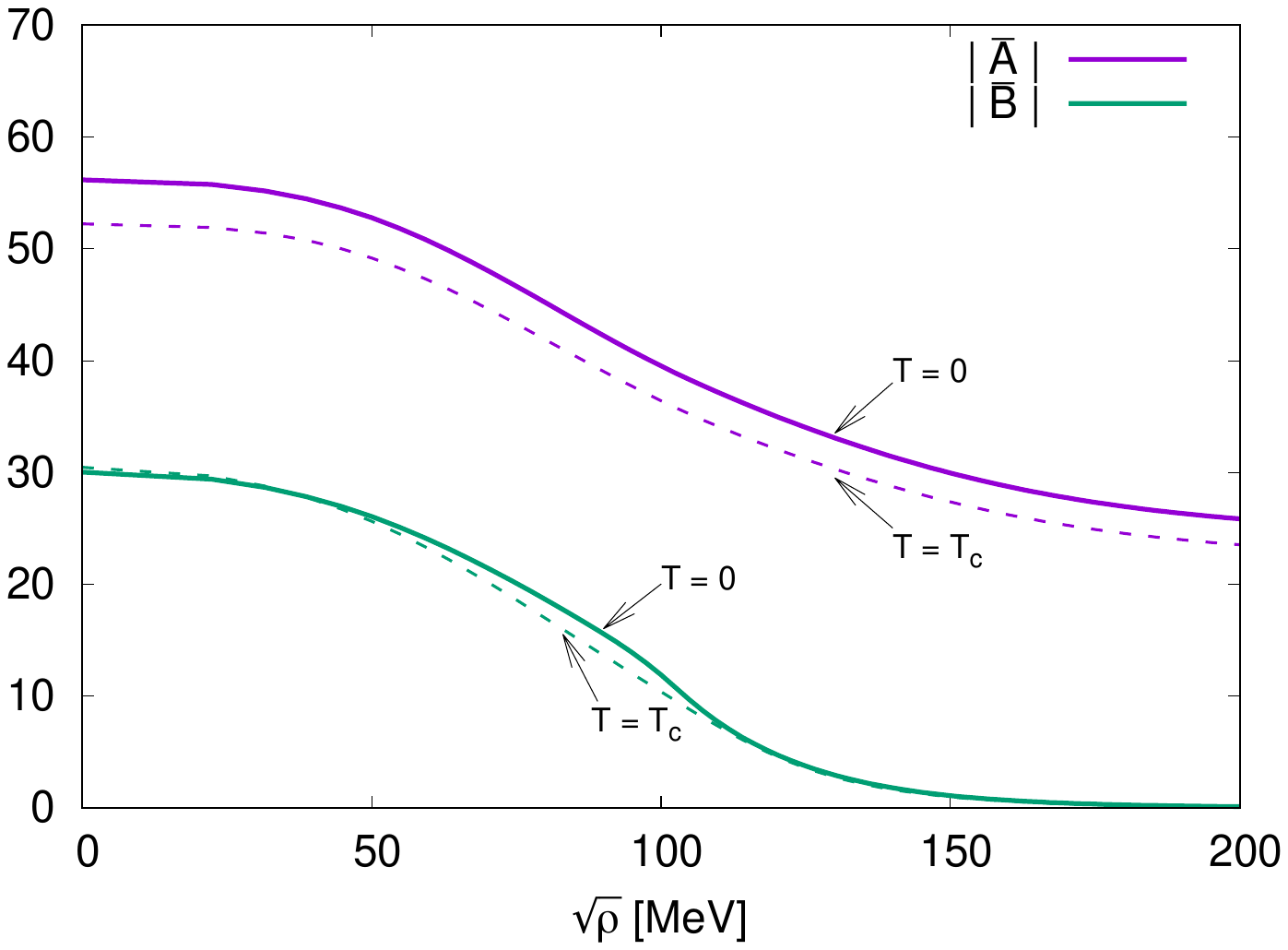}
\caption{Dimensionless anomaly coefficient functions at $T=0$ and $T=T_c$, as a function of the chirally invariant combination $\rho$, for a $T$-independent initial anomaly parameter. Both functions monotonically decrease showing that as the condensates evaporate, fluctuations tend to strengthen the axial anomaly. The scale is set to $k/\Lambda=0.1$.}
\label{Fig:ABfunc}
\end{figure}

\begin{figure}[]
\includegraphics[bb = 350 80 495 590,scale=0.33]{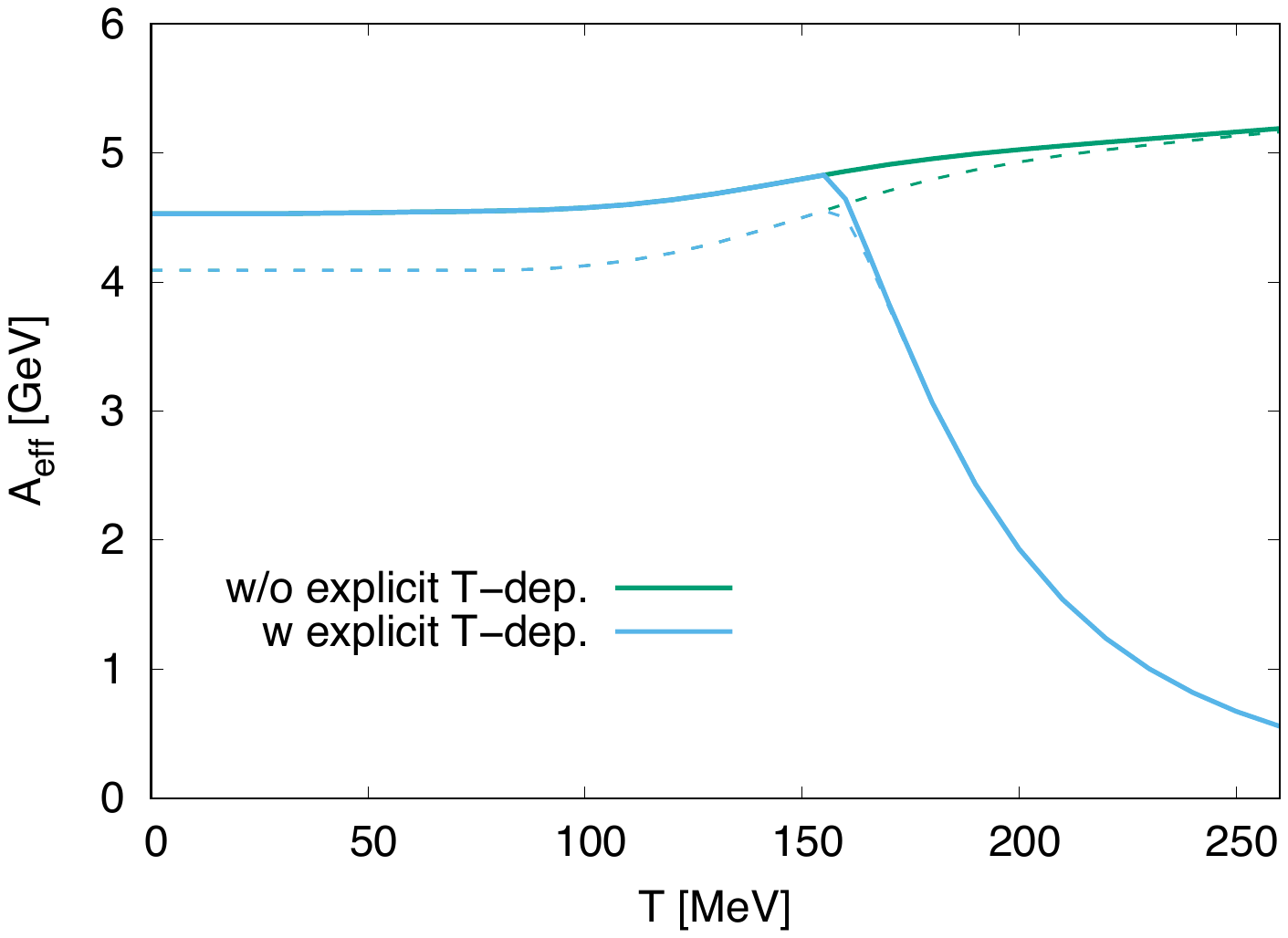}
\caption{Strength of the effective anomaly coupling, $A_{\eff}(T)=A(\rho_{\min}(T))+B(\rho_{\min}(T))\Delta_{\min}(T)$, as a function of the temperature. Dashed lines are only indicated for reference, those curves include only $A$ (and not $B$) in the effective coupling.}
\label{Fig:anomT}
\end{figure}

Now let us focus on the fate of the axial anomaly as the function of the temperature. In Fig. \ref{Fig:ABfunc} we show, for a temperature independent initial $a$ anomaly parameter, the absolute value of the dimensionless, $\bar{A}(\rho)=A(\rho)/k$, $\bar{B}(\rho)=B(\rho)k^2$ anomaly coefficient functions (note that both $A$ and $B$ are {\it negative}). First, we note that, the shape of each function does not depend very strongly on the temperature, but the actual anomaly strength does,  as the evaporation of $v_{\ns}$ and $v_{\s}$ yields the chirally invariant combination $\rho_{min}$ to show a decreasing tendency with $T$. This is in agreement with our earlier findings \cite{fejos16,fejos21} showing that mesonic fluctuation effects strengthens the anomaly. Now we see that on top of the $|Q|=1$ sector, the same applies for interactions coming from instantons with $|Q|=2$ topological charge. Second, Fig. \ref{Fig:ABfunc} also tells us that in order to reproduce the correct high-$T$ behavior of the axial anomaly, we absolutely must include the damping effects through the explicit temperature dependence of the initial anomaly coefficient, as already announced in the previous paragraph. This is illustrated in Fig. \ref{Fig:anomT}, where the effective anomaly strength, 
\bea
A_{\eff}(T) = A(\rho_{\min}(T))+B(\rho_{\min}(T))\Delta_{\min}(T)
\eea
is plotted against the temperature.  The ``min'' subscripts refer to the actual minimum point of the complete effective potential, that is where the effective interaction is defined. We see that by neglecting the explicit temperature dependence of the initial $a$ anomaly parameter, $A_{\eff}$ would monotonically increase with $T$ even beyond $T_c$. To avoid such a scenario did we employ the assumption (\ref{Eq:aT}). As a result, we still see an intermediate strengthening of the $U_A(1)$ anomaly toward $T\rightarrow T_c$, however, including the $|Q|=2$ interactions somewhat moderates the effect compared to our earlier findings, which included only the $|Q|=1$ sector \cite{fejos21}.

\begin{figure}[t]
\includegraphics[bb = 350 80 495 590,scale=0.33]{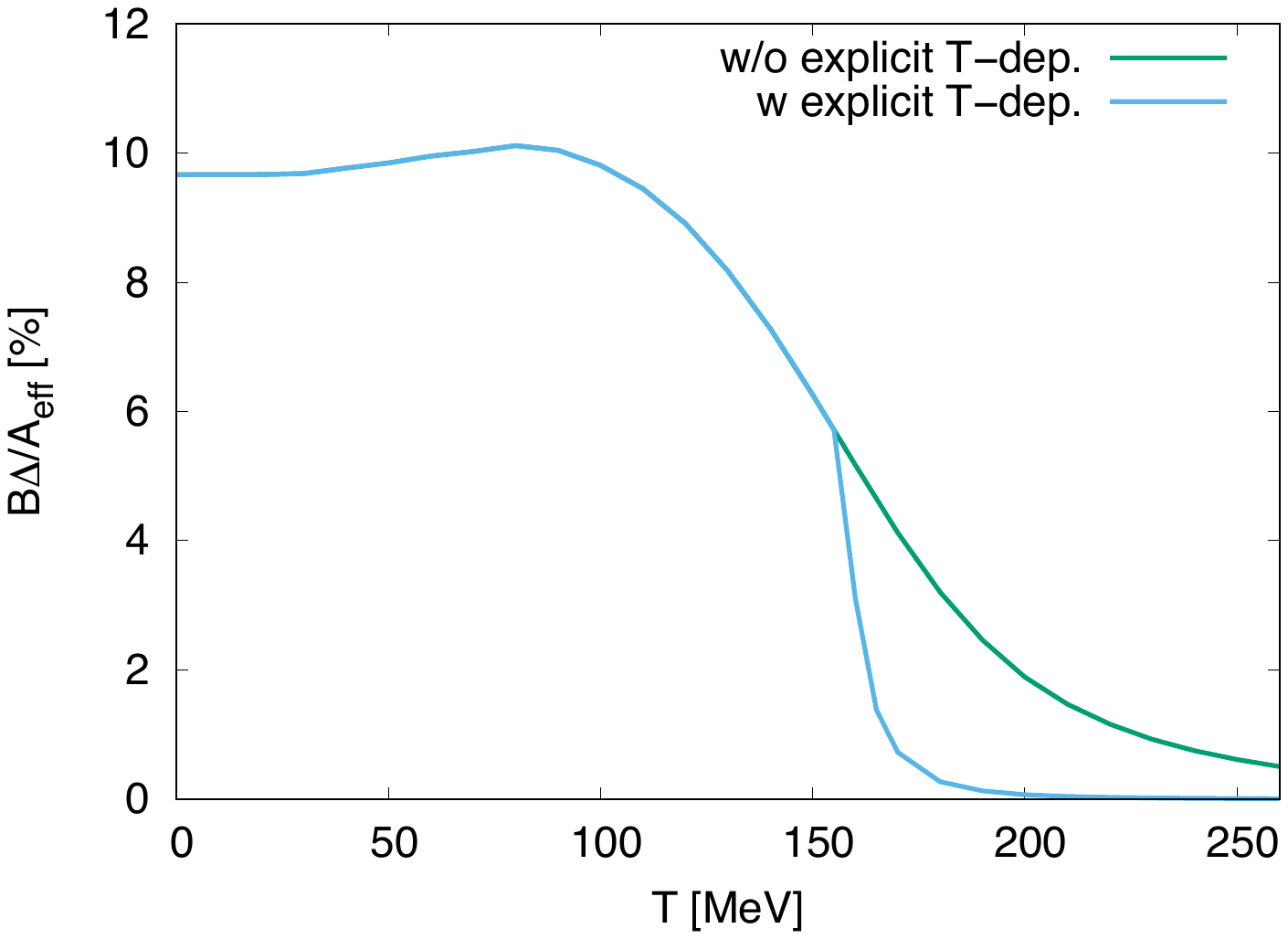}
\caption{Importance of the $|Q|=2$ contribution to the effective anomaly coupling for both scenarios. Ratio of the $|Q|=2$ contribution to the full effective anomaly coupling is plotted.}
\label{Fig:anomratio}
\end{figure}

\begin{figure}
\includegraphics[bb = 350 80 495 590,scale=0.33]{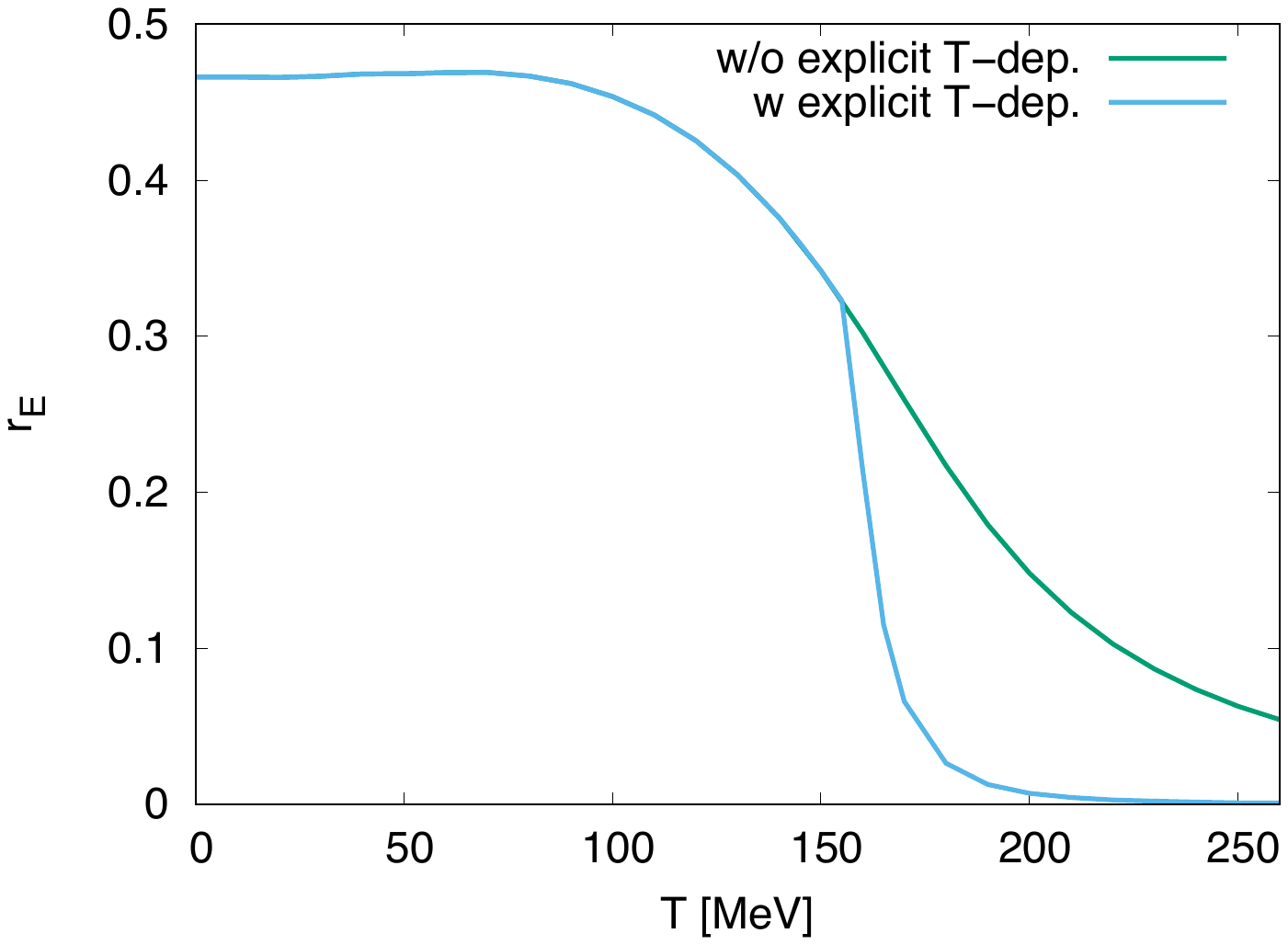}
\caption{Ratio between the chirally symmetric and anomalous parts of the effective potential as a function of the temperature.}
\label{Fig:enratio}
\end{figure}

One may now wonder about the importance of the $|Q|=2$ interactions, and this is what is shown in Fig. \ref{Fig:anomratio}. We plot the ratio between the $|Q|=2$ contribution to the effective anomaly coupling and the latter itself. Interestingly, for this quantity both discussed scenarios show quite a rapid decrease (obviously the one that corresponds to an initially temperature dependent anomaly coupling is faster), but the most important observation is that for $T \lesssim T_c$ the $|Q|=2$ topological sector provides around $10\%$ of the anomaly strength. This confirms that for $T \lesssim T_c$, instanton interactions with $|Q|=2$ topological winding are definitely important, and we also see, however, that, for $T \gtrsim T_c$ it is safe to keep only the $|Q|=1$ sector, as expected from a dilute instanton gas approximation in the underlying theory. Finally, we define
\bea
r_E&=&\frac{|A(\rho_{\min})\Delta_{\min}+B(\rho_{\min})\Delta_{\min}^2|}{U(\rho_{\min})+C(\rho_{\min})\tau_{\min}} \nonumber\\
&\equiv& \frac{|A_{\eff}\Delta_{\min}|}{U(\rho_{\min})+C(\rho_{\min})\tau_{\min}},
\eea
as the absolute value of the ratio between the chirally symmetric and anomalous contributions in the effective potential\footnote{We note that $U(\rho)$ is normalized such that $U(0)=0$.}. Interestingly, as it can be seen in Fig. \ref{Fig:enratio}, for lower temperatures they are almost of the same order, but for $T\gtrsim T_c$, the anomalous contributions rapidly disappear, much faster than the effective anomaly coupling itself.  This is easily understood, since the numerator of $r_E$ contains an extra factor of $v_{\ns}^2$ evaporating fast at high $T$, compared to $A_{\eff}$. This comes from the cubic invariant being $\Delta \sim v_{\s}v_{\ns}^2$. 

Before moving on to the mass spectra we note that in \cite{pisarski24} it is not entirely ruled out that in the chiral limit, at the critical point all anomaly couplings may simultaneously vanish. In the light of the present results this would mean a rather radical change in the behavior of the topological fluctuations compared to those at the physical point.

\begin{figure}[t]
\includegraphics[bb = 350 80 495 590,scale=0.33]{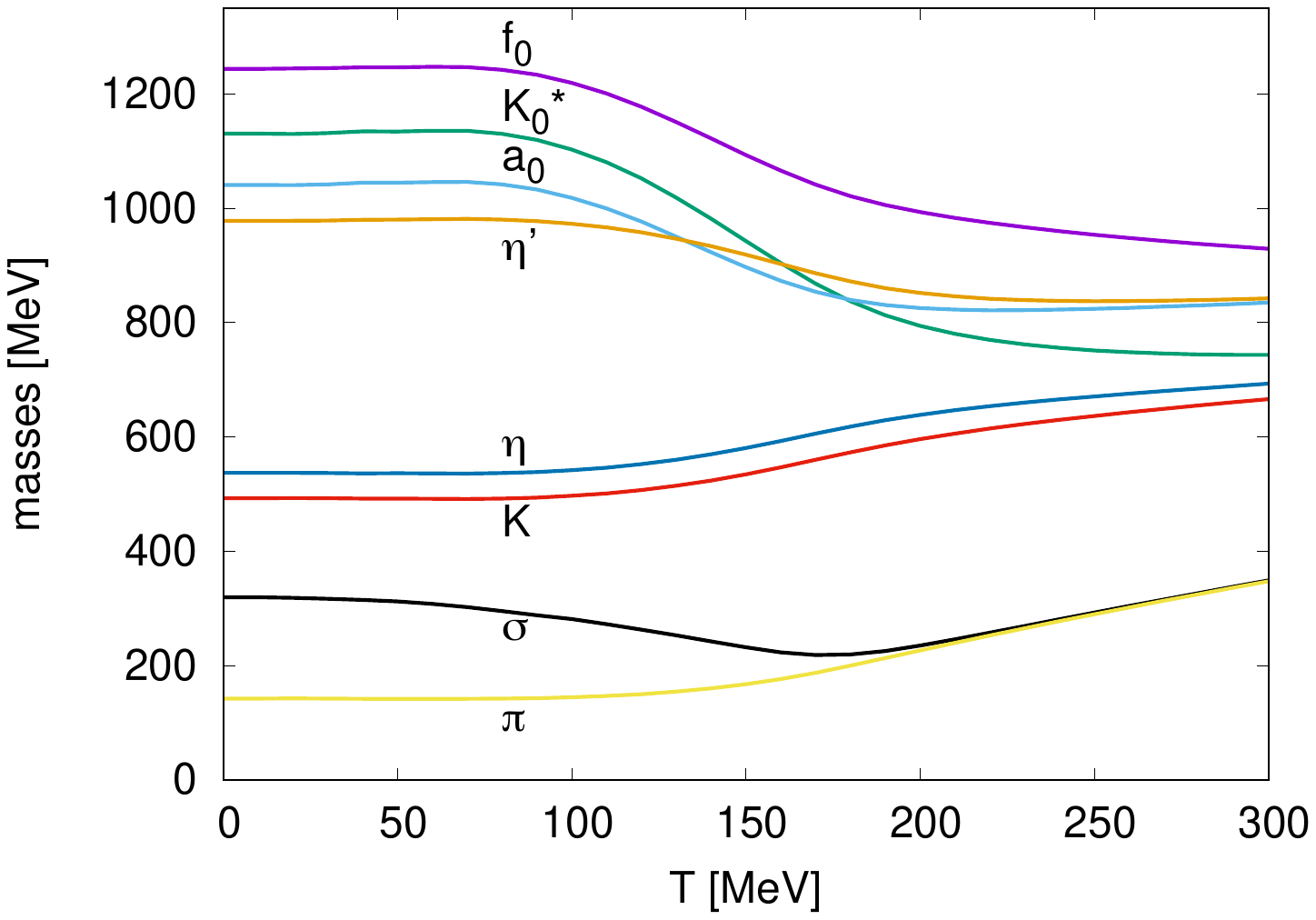}
\caption{Thermal behavior of the mesonic spectrum with a temperature {\it independent} initial anomaly parameter. }
\label{Fig:masses1}
\includegraphics[bb = 350 80 495 590,scale=0.33]{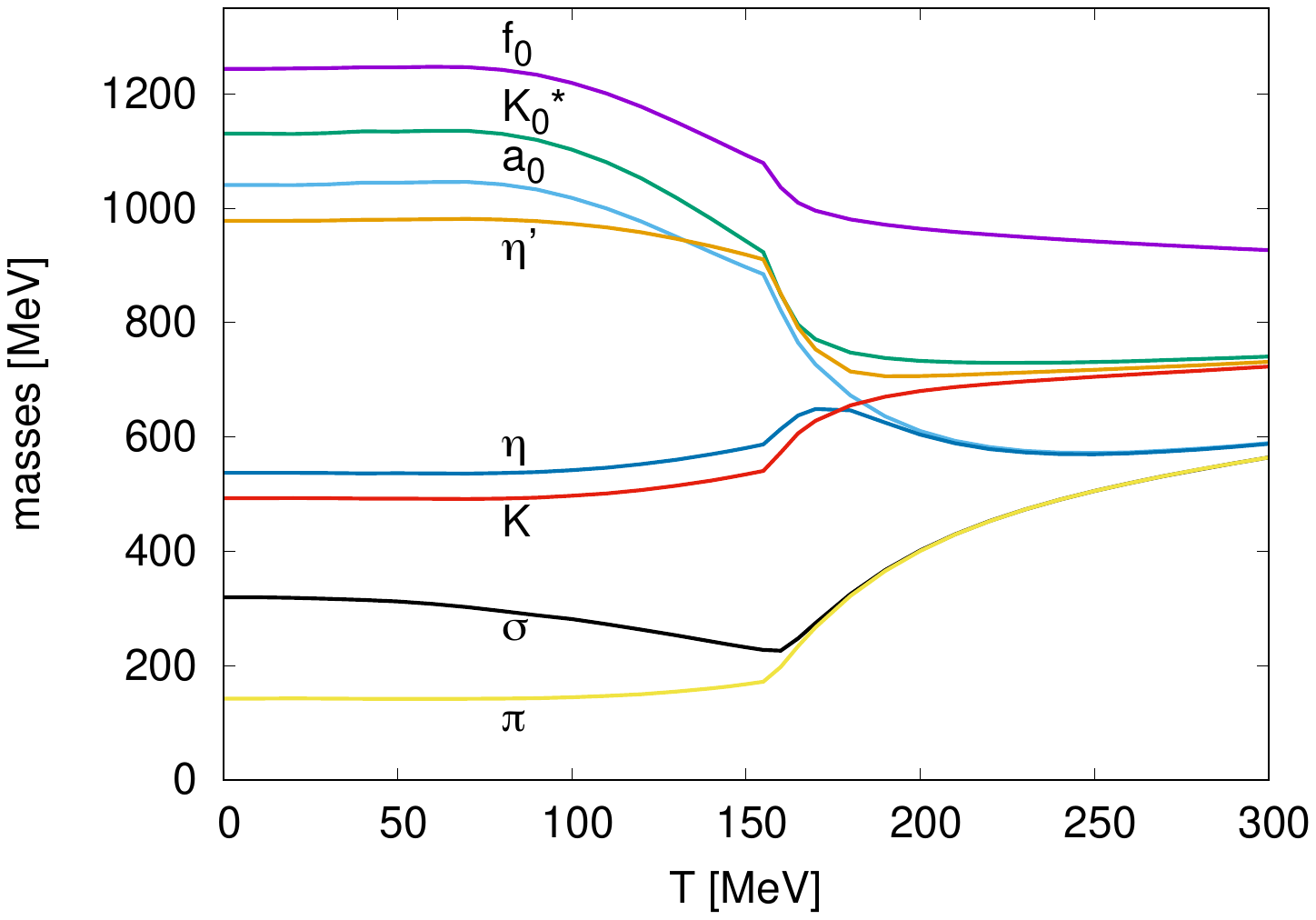}
\caption{Thermal behavior of the mesonic spectrum with a temperature {\it dependent} initial anomaly parameter. }
\label{Fig:masses2}
\end{figure} 

\subsection{Mass spectrum}

The mass spectrum is obtained by diagonalizing the second derivative tensor of $V_{k=0}$ [see (\ref{Eq:Vfinal})] at the minimum point of the complete effective potential, computed in the background of the physical condensates $v_0(T),v_8(T)$ [or, equivalently, $v_{\ns}(T)$ and $v_{\s}(T)$]. For this we have to obtain numerically the first and second derivatives of the $U(\rho)$, $C(\rho)$, $A(\rho)$, and $B(\rho)$ functions, and make use of the derivatives of the $\rho$, $\tau$ and $\Delta$ invariants with respect to all field variables. Analytical formulas for the latter can be found in \cite{fejos21}. 

The $T$-dependence of the spectrum for our two scenarios can be seen in Fig. \ref{Fig:masses1} and Fig. \ref{Fig:masses2}, respectively. One notices the different high temperature behavior for each case.  In Fig. \ref{Fig:masses1}, at high $T$ the anomaly coupling $A_{\eff}$ stays nonzero, though among the condensates, $v_{\ns}$ practically disappears. The strange condensate $v_{\s}$ is still significant. As for Fig. \ref{Fig:masses2}, $A_{\eff}$ is damped according to (\ref{Eq:aT}) and it practically vanishes together with $v_{\ns}$, but $v_{\s}$ is still alive.  In what follows, we provide a semi quantitative analysis of the observed high temperature spectra based on these remarks.

The basic multiplet structure is the same for both cases in the two component $(v_{\ns},v_{\s})$ background. We have both in the pseudoscalar and scalar sectors one triplet ($\pi$ and $a_0$), one quartet ($K$ and $K_0^*$), and the mixing $(\eta,\eta^\prime)$ pair in the pseudoscalar, and $(\sigma,f_0)$ in the scalar sector, respectively. We do not list here exact but lengthy mass formulas separately, instead restrict ourselves on the presentation of useful relations, which can clarify the behaviors observed in Fig. \ref{Fig:masses1} and Fig. \ref{Fig:masses2},  reflecting especially on the survival/suppression of the anomaly coupling.  For the sake of transparency and simplicity, in the forthcoming analysis we neglect the $\rho$ dependence of the $A(\rho)$, $B(\rho)$ and $C(\rho)$ coefficient functions, and restrict ourselves to $U(\rho)=m^2\rho+\lambda_1\rho^2$. Still, one can nicely interpret some of the characteristic features of the observed high-$T$ spectra.

First, we have two compact expressions for the scalar-pseudoscalar difference in the triplet and the quartet sectors:
\begin{subequations}
\bea
M^2_{a_0}-M^2_\pi&=&-\sqrt{2}(A+2B\Delta)v_{\s}+Cv_{\ns}^2,\\
M^2_{K_0^*}-M^2_K&=&-(A+2B\Delta)v_{\ns}+\sqrt{2}Cv_{\s}v_{\ns}.
\eea
\end{subequations}
As we saw already, near and above the pseudocritical temperature, the nonstrange condensate quickly diminishes, while $v_s$ stays nearly constant. This means that in the triplet sector a nonzero mass difference in the high temperature region signals the persistence of the anomaly. 

Below, in the high temperature region we set $v_{\ns}\approx 0$, and then even the expressions of the actual masses become rather simple (see also \cite{meggiolaro23}):
\begin{subequations}
\bea
M^2_\pi&=&m^2+\lambda_1 v_{\s}^2-\frac{1}{3}C v_{\s}^2+\frac{1}{\sqrt{2}}Av_{\s},\\
M^2_{a_0}&=&m^2+\lambda_1 v_{\s}^2-\frac{1}{3}Cv_{\s}^2-\frac{1}{\sqrt{2}}Av_{\s},\\
M^2_K&=&M^2_{K_0^*}= m^2+\lambda_1 v_{\s}^2+\frac{2}{3}Cv_{\s}^2.
\eea 
\end{subequations}
The ${\cal D}$ discriminant constructed from the eigenvalue equation of the mixing matrix in the pseudoscalar $(0-8)$ sector determines the $\eta-\eta^\prime$ mass difference, which, for a vanishing nonstrange condensate is rather transparent:
\bea
M^2_{\eta^\prime}-M^2_\eta \equiv \sqrt{{\cal D}}=\frac{1}{\sqrt2}Av_{\s}+Cv_{\s}^2.
\eea
This difference and also the above mass formulas can be discussed assuming two limiting situations, depending on the relative magnitude of the mass contribution from the nonanomalous and anomalous parts of the potential. When $Cv_s^2\gg-Av_s$, which is certainly true when $A$ is suppressed by its explicit $T$-dependence, one finds asymptotically
\bea
M^2_{\eta^\prime}= m^2+\lambda_1v_{\s}^2+\frac{2}{3}Cv_s^2,\quad M^2_{\eta}=m^2+\lambda_1v_{\s}^2-\frac{1}{3}Cv_s^2.\nonumber\\
\eea
This means that in this case the levels of $M^2_{a_0},M^2_\pi,M^2_\eta$ and $M^2_{K_0^*},M^2_K,M^2_{\eta^\prime}$ group separately. This can be clearly observed in Fig. \ref{Fig:masses2}.

 On the other hand, when the anomaly survives and dominates, i.e. $-Av_s \gg Cv_s^2$, one finds
\bea
\!\!\!\!\!M^2_{\eta^\prime}=m^2+\lambda_1 v^2_{\s}-\frac{1}{\sqrt{2}}Av_{\s}, \quad M^2_\eta= m^2+\lambda_1 v^2_{\s}.
\eea
In this case the lighter excitation of the pseudoscalar $(0-8)$ sector, i.e., $\eta$, becomes nearly degenerate with $K$ and $K_0^*$, while the heavier $\eta'$ will be located close to $a_0$. Note that, the anomaly sustains a mass difference between $\pi$ and $a_0$. This characteristics is observed in Fig. \ref{Fig:masses1}, where one assumed no explicit temperature dependence for the initial anomaly coefficient $a$. 

Based on the above analysis, by observing the variation of the meson masses as the temperature is gradually increased, such rearrangement of $\eta$ and $\eta^\prime$ would clearly indicate the onset of anomaly suppression.  We also note that the current setup shows that the drop in the $\eta'$ mass at $T_c$ becomes larger compared to our earlier results, hinting that if this is maintained at the partial recovery of chiral symmetry at the nuclear liquid-gas transition, the $\eta'$-nucleon bound state might indeed be observable in a nuclear medium \cite{jido19,sakai23}.

Also, we note that a similar analysis in the scalar $(0-8)$ sector shows that, the $\sigma$ and $f_0$ masses in the $v_{\ns}\approx 0$ limit become
\begin{subequations}
\bea
M_{\sigma}^2&=&m^2+\lambda_1 v_{\s}^2-\frac13 Cv_{\s}^2+\frac{1}{\sqrt2}A v_{\s}, \\
M_{f_0}^2&=&m^2+3\lambda_1 v_{\s}^2+2Cv_{\s}^2.
\eea
\end{subequations}
Both scenarios $Cv_s^2\gg-Av_s$ and $-Av_s\gg Cv_s^2$ show that $\sigma$ degenerates with $\pi$ at high $T$, and it is also revealed that the $f_0$ mass is insensitive to the anomaly even when the latter dominates.

Finally we comment on the mixing angles in the scalar and pseudoscalar sectors. Note that, at first, one usually defines the scalar and pseudoscalar mixing angles, $\Theta_{s/\pi}$, in the $0-8$ basis,  which are then transformed into $\varphi_{s/\pi}$, corresponding to the ns $-$ s coordinate system, through the ideal mixing matrix (see, e.g., the Appendix of \cite{rennecke16}). Depending on the orientation (i.e. the sign) of $\Theta_{s/\pi}$, one ends up with curves for the temperature dependence of $\varphi_{s/\pi}$ that are similar to either \cite{kovacs16} or \cite{rennecke16}. Since our model and method are closer to that of \cite{rennecke16}, we adopt their convention, which, in the pseudoscalar sector leads to
\begin{subequations}
\bea
\sin 2\Theta_{\pi} &=& -\frac{2M_{\pi,08}^2}{M_{\eta'}^2-M_{\eta}^2}, \\
\cos 2\Theta_{\pi} &=& \frac{2M_{\pi,88}^2-M_{\eta'}^2-M_{\eta}^2}{M_{\eta'}^2-M_{\eta}^2}.
\eea
\end{subequations}
The second formula is especially convenient for calculating the angle, as it is not sensitive to either its orientation, or a possible level crossing between $M_{\pi,00}^2$ and $M_{\pi,88}^2$. As the numerics show that $M_{\pi,08}^2<0$, in the above convention both $\Theta_{\pi}>0$ and $\varphi_{\pi}>0$. Our results are very similar to that of Fig. 8 in \cite{rennecke16}. We see an almost completely identical behavior for $\varphi_s$, but more interestingly, temperature dependence of $\varphi_{\pi}$ via our $a(T)$ ansatz is very close to that of the LPA'+Y approximation of \cite{rennecke16}, showing that $\varphi_{\pi}$ goes to zero at high temperature. It seems that the effect of the wave function renormalization is similar to a temperature dependent bare anomaly parameter, introduced in (\ref{Eq:aT}).

\section{Summary}

In this paper we analyzed the thermal behavior of the axial anomaly using the functional renormalization group method. Compared to our earlier study \cite{fejos21}, we applied an enlarged set of terms in the effective potential of the $N_f=3$ meson model. We derived scale evolution equations for the effective potential that include interactions originated from instantons of the underlying theory of QCD with arbitrary topological charge [see Eq. (\ref{Eq:flowU})].  We discussed two distinct scenarios: (1) the initial $|Q|=1$ anomaly coupling is an environment independent constant, and (2) the initial $|Q|=1$ anomaly coupling inherits explicit temperature dependence from QCD interactions occurring above the initial momentum scale.

We numerically solved the flow equations for the chiral condensate dependent anomaly couplings coming from instantons with $|Q|=1,2$ topological charges and found that for low temperatures the $|Q|=2$ interaction contributes around $10\%$ to the effective $U_A(1)$ coupling $A_{\eff}$. The ratio between the $|Q|=2$ and $|Q|=1$ contributions to $A_{\eff}$ rapidly decreases once the temperature hits $T_c$, showing that at high temperatures it is sufficient to include instanton interactions with $|Q|=1$. For $T\lesssim T_c$, however, $|Q|=2$ interactions cannot be dropped.

In accordance with our earlier findings we saw that mesonic fluctuations tend to strengthen the anomaly couplings toward the critical point, which are then suppressed for $T\gtrsim T_c$ by instanton effects. We also calculated numerically the mass spectrum and provided semiquantitative analytical formulas to understand its high temperature behavior.  In particular, we showed that the $\eta$ and $\eta'$ mass difference, on top of the anomalous terms, also contains one of the quartic couplings through the strange condensate. As a result, we argued that the $\eta$ and $\eta'$ clustering with other excitations occur differently for weak (temperature suppressed) and strong (temperature sustained) anomaly. Observing the finite temperature spectrum in lattice simulations may, therefore, hint on the actual thermal behavior of the $U_A(1)$ anomaly coupling.

Our analysis could be extended in a straightforward fashion for couplings that correspond to $|Q|>2$ instanton interactions by expanding (\ref{Eq:flowU}) to higher order in $\Delta$. Our method also offers the possibility of treating all anomalous interactions on an equal footing by solving the flow equation for $U(\rho,\Delta$) on a two dimensional grid. Though numerically this might be very demanding, it would certainly provide the most general answer to the question of the importance of all anomalous interactions. Furthermore, it could be interesting to apply our method to two-color QCD with $N_f=2$, where the global symmetry of the system turns into $SU(4)$ (Pauli-Gursey symmetry) \cite{kawaguchi23}. The technique presented here, equipped with the corresponding lattice data could provide new insight into the behavior of the anomaly in such a system.

\section*{Acknowledgments}

This research was supported by the Hungarian National Research, Development, and Innovation Fund under Project No. FK142594. G.F. also received support from the Hungarian State Eötvös Scholarship. The authors would like to express their gratitude to Naoki Yamamoto for his insights regarding the explicit temperature dependence of the initial anomaly parameter in the ultraviolet.

\vspace{0.2cm}
\makeatletter
\@addtoreset{equation}{section}
\makeatother 
\renewcommand{\theequation}{A\arabic{equation}} 

\appendix

\section*{Appendix: Flow equations}
\setcounter{equation}{0} 

Here we list the explicit flow equations that were solved numerically. The flow equation for $U_k(\rho,\Delta)$, determined in Sec. III, using the $\langle M \rangle = (s_0+i\pi_0)T_0$ background is the following:
\begin{widetext}
\bea
\label{Eq:flowU}
\partial_k U_k(\rho,\Delta)&=& \Omega_d \frac{4k^d}{d}T\sum_{n}\tilde{\partial}_k \log\left((\omega_n^2+k^2+U_{k,\rho})^2+\frac{4}{3}(\omega_n^2+k^2+U_{k,\rho})\rho C_k-\frac{1}{3}\rho U_{k,\Delta}^2+2\Delta C_kU_{k,\Delta}\right)\nonumber\\
&+&\Omega_d \frac{k^d}{2d}T \sum_{n}\tilde{\partial}_k\log\Biggl[ (\omega_n^2 + k^2+U_{k,\rho}+3\Delta U_{k,\rho\Delta})^2+(\omega_n^2+k^2+U_{k,\rho})\left(2\rho U_{k,\rho\rho}+\frac{2}{3}\rho^2U_{k,\Delta\Delta}\right)\nonumber\\
&&\hspace{2.6cm}-6\Delta U_{k,\Delta}U_{k,\rho\rho}-\frac{4}{3}\rho U_{k,\Delta}(U_{k,\Delta}+2\rho U_{k,\rho\Delta})-\frac{4}{3}\rho^3U_{k,\rho\Delta}^2 \nonumber\\
&&\hspace{2.6cm}-U_{k,\Delta\Delta}\left(2\rho\Delta U_{k,\Delta}-\frac{1}{3}(4\rho^3-27\Delta^2)U_{k,\rho\rho}\right)\Biggr],
\eea
where the subscripts $\rho$ and/or $\Delta$ refer to partial derivatives with respect to the corresponding variable. The flow equation for $C_k(\rho)$, which was calculated in the purely imaginary 
$\langle M \rangle = i(\pi_0T_0+\pi_8T_8)$ background, takes the form of
\bea
\partial_k C_k(\rho)&=& \Omega_d \frac{k^d}{d}T\sum_{n}\tilde{\partial}_k \Biggl\{\frac{U_{k,\Delta}^2}{2\rho}\left(\frac{1}{D_0}-\frac{1}{2D_8}-\frac{(\omega_n^2+k^2+U_{k,\rho})^2}{2D_0D_8}\right)\nonumber\\
&&\hspace{2.3cm}+\frac{3}{4\rho}(\omega_n^2+k^2+U_{k,\rho})\left(U_{k,\rho\rho}-\frac{2}{3}C_k\right)\left(\frac{1}{D_8}-\frac{1}{D_0}\right)\nonumber\\
&&\hspace{2.3cm}+\frac{1}{2D_0}\left[\rho U_{k,\Delta\Delta}\left(\frac{2}{3}C_k-U_{k,\rho\rho}\right)+U_{k,\rho\Delta}(2U_{k,\Delta}+\rho U_{k,\rho\Delta})\right]\nonumber\\
&&\hspace{0.6cm}-\frac{1}{2D_0D_8}\Biggl[(\omega_n^2+k^2+U_{k,\rho})\left(2U_{k,\rho\rho}\left(\frac{1}{2}U_{k,\Delta}-\rho U_{k,\rho\Delta}\right)^2+\frac{2}{3}(2C_k+\rho U_{k,\Delta\Delta})\left(\frac{1}{2}U_{k,\Delta}+\rho U_{k,\rho\Delta}\right)^2\right)\nonumber\\
&&\hspace{3.0cm}+2(\omega_n^2+k^2+U_{k,\rho})^2\rho U_{k,\rho\Delta}^2+\frac{2}{9}\rho^2U_{k,\Delta\Delta}C(U_{k,\Delta}+2\rho U_{k,\rho\Delta})^2\Biggr]\Biggr\}\nonumber
\eea
\bea
&&\hspace{1cm}-\Omega_d \frac{k^d}{2d}T\sum_{n}\tilde{\partial}_k\Biggl\{\frac{1}{3D_0D_8}\Biggl[(\omega_n^2+k^2+U_{k,\rho}+2\rho U_{k,\rho\rho})\left(\omega_n^2+k^2+U_{k,\rho}+\frac{4}{3}\rho C\right)(2C_k-\rho U_{k,\Delta\Delta})^2\nonumber\\
&&\hspace{3.7cm}+\left(\omega_n^2+k^2+U_{k,\rho}+\frac{2}{3}\rho^2 U_{k,\Delta\Delta}\right)(\omega_n^2+k^2+U_{k,\rho})(3U_{k,\rho\rho}+4(C_k+\rho C_{k,\rho}))^2\Biggr]\nonumber\\
&&\hspace{2.5cm}-\frac{1}{3D_0D_8}(3U_{k,\rho\rho}+4(C_k+\rho C_{k,\rho}))\Bigl[U_{k,\Delta}\left(\omega_n^2+k^2+U_{k,\rho}+\frac{2}{3}\rho^2U_{k,\Delta\Delta}\right)(2\rho U_{k,\rho\Delta}+U_{k,\Delta})\nonumber\\
&&\hspace{7.8cm}+2(U_{k,\Delta}+\rho U_{k,\rho\Delta})(\omega_n^2+k^2+U_{k,\rho})(2\rho U_{k,\rho\Delta}-U_{k,\Delta})\Bigr]\nonumber\\
&&\hspace{2.5cm}+\frac{1}{3D_0D_8}(2C_k-\rho U_{k,\Delta\Delta})\Bigl[(\omega_n^2+k^2+U_{k,\rho}+2\rho U_{k,\rho\rho})U_{k,\Delta}(2\rho U_{k,\rho\Delta}-U_{k,\Delta})\nonumber\\
&&\hspace{6.5cm}+2\big(\omega_n^2+k^2+U_{k,\rho}+\frac{4}{3}\rho C\big)(U_{k,\Delta}+\rho U_{k,\rho\Delta})(2\rho U_{k,\rho\Delta}+U_{k,\Delta})\Bigr] \nonumber\\
&&\hspace{1cm}-\frac{4}{3D_0D_8}\left[\frac{1}{4}U_{k,\Delta}^2-\rho^2U_{k,\rho\Delta}^2+\rho(2C_k-\rho U_{k,\Delta\Delta})\left(U_{k,\rho\rho}+\frac{4}{3}(C_k+\rho C_{k,\rho})\right)\right]U_{k,\Delta}(U_{k,\Delta}+\rho U_{k,\rho\Delta})\Biggr\}\nonumber
\eea
\bea
\label{Eq:flowC}
&&\hspace{1cm}+\Omega_d \frac{k^d}{d}T\sum_{n}\tilde{\partial}_k\Biggl\{\Big[\frac{7}{D_8}\left(C_k^2+(\omega_n^2+k^2+U_{k,\rho})C_{k,\rho}+\frac{2}{3}\rho C_kC_{k,\rho}\right)\nonumber\\
&&\hspace{2.6cm}-\frac{8}{3D_8^2}\left(\frac{1}{2}U_{k,\Delta}^2+\frac{2}{3}\rho C_k^2+2C(\omega_n^2+k^2+U_{k,\rho})\right)^2\Big]\nonumber\\
&&\hspace{1.9cm}+\frac{1}{2D_0}\Biggl[\left(C_{k,\rho}-\frac{3}{2}U_{k,\Delta\Delta}\right)(\omega_n^2+k^2+U_{k,\rho}+2\rho U_{k,\rho\rho})\nonumber\\
&&\hspace{2.9cm}+(5C_{k,\rho}+2\rho C_{k,\rho\rho})\left(\omega_n^2+k^2+U_{k,\rho}+\frac{2}{3}\rho^2U_{k,\Delta\Delta}\right)+3U_{k,\rho\Delta}(U_{k,\Delta}+\rho U_{k,\rho\Delta})\Biggr]\nonumber\\
&&\hspace{1.9cm}+\frac{1}{2D_8}\Biggl[(\omega_n^2+k^2+U_ {k,\rho})\left(6C_{k,\rho}+\frac{1}{2}U_{k,\Delta\Delta}\right)-U_{k,\Delta}U_{k,\rho\Delta}+2C_k^2+\frac{4}{3}\rho C_k\left(C_{k,\rho}+\frac{1}{2}U_{k,\Delta\Delta}\right)\nonumber\\
&&\hspace{1.9cm}-\frac{1}{3D_8}\left[4C_k(\omega_n^2+k^2+U_{k,\rho})+U_{k,\Delta}^2+\frac{4}{3}\rho C_k^2\right]^2\Biggr]\Biggr\},
\eea
where
\begin{subequations}
\bea
D_0&=&(\omega_n^2+k^2+U_{k,\rho}+2\rho U_{k,\rho\rho})\left(\omega_n^2+k^2+U_{k,\rho}+\frac{2}{3}\rho^2U_{k,\Delta\Delta}\right)-\frac{4}{3}\rho (U_{k,\Delta}+\rho U_{k,\rho\Delta})^2, \\
D_8&=&\left(\omega_n^2+k^2+U_{k,\rho}\right)\left(\omega_n^2+k^2+U_{k,\rho}+\frac{4}{3}\rho C_k\right)-\frac{1}{3}\rho U_{k,\Delta}^2.
\eea
\end{subequations}
\end{widetext}


\begin{thebibliography}{9}

\bibitem{schaefer96}T. Schaefer, Phys. Lett. B{\bf 389}, 445 (1996).
\bibitem{schaefer98}T. Schaefer and E. Shuryak, Rev. Mod. Phys. {\bf 70}, 323 (1998).
\bibitem{ding21a} H.-T. Ding, W.-P. Huang, M. Lin, S. Mukherjee, P. Petreczky and Y. Zhang, Proc. Sci. LATTICE2021 ({\bf 2022}) 591 [arXiv:2112.00318].
\bibitem{vig21} R. A. Vig and T. G. Kovacs, Phys. Rev. D{\bf 103}, 114510 (2021).
\bibitem{shuryak82} E. V. Shuryak, Nucl. Phys. {\bf B203}, 116 (1982).
\bibitem{kaczmarek20} O. Kaczmarek, F. Karsch, A. Lahiri, L. Mazur, and C. Schmidt, NIC series in {\it Proceedings of 10th NIC Symposium} (2020), p. 193, https://juser.fz-juelich.de/record/874262.
\bibitem{buchoff14} M.I. Buchoff, M. Cheng, N.H. Christ, H.T. Ding, C. Jung, F. Karsch {\it et al.}, Phys. Rev. D{\bf 89}, 054514 (2014).
\bibitem{bhattacharya14} T. Bhattacharya, M.I. Buchoff, N.H. Christ, H.T. Ding, R. Gupta, C. Jung {\it et al.}, Phys. Rev. Lett. {\bf 113},  082001 (2014).
\bibitem{bazazov12} A. Bazazov, T. Bhattacharya, M. Cheng, C. DeTar, H. T. Ding, S. Gottlieb et al.,  Phys. Rev. D{\bf 85}, 054503 (2012).
\bibitem{brandt16} B. Brandt, A. Francis, H. B. Meyer, O. Philipsen, D. Robaina, and H. Wittig, J. High Energy Phys. 12 (2016) 158.
\bibitem{dick15} V. Dick, F. Karsch, E. Laermann, S. Mukherjee, and S. Sharma, Phys. Rev. D{\bf 91},  094504 (2015).
\bibitem{ding21} H. T. Ding, S.T. Li, S. Mukherjee, A. Tomiya, X.D. Wang, and Y. Zhang, Phys. Rev. Lett. {\bf 126}, 082001 (2021).
\bibitem{kaczmarek21} O. Kaczmarek, L. Mazur, and S. Sharma, Phys. Rev. D{\bf 104}, 094518 (2021). 
\bibitem{tomiya16} A. Tomiya, G. Cossu, S. Aoki, H. Fukaya, S. Hashimoto, T. Kaneko, and J. Noaki, Phys. Rev. D{\bf 96}, 034509 (2017).
\bibitem{aoki21} S. Aoki, Y. Aoki, H. Fukaya, S. Hashimoto, C. Rohrhofer, and K. Suzuki,  Prog. Theor. Exp. Phys. {\bf 2022}, 023B05 (2022).
\bibitem{borsanyi21} Sz. Borsanyi and D. Sexty, Phys. Lett. B{\bf 815}, 136148 (2021).
\bibitem{lahiri21} A. Lahiri, Proc. Sci. LATTICE2021 (2022) 003 [arXiv:2112.08164].
\bibitem{kovacs16} P. Kovacs, Zs. Szep, and Gy. Wolf, Phys. Rev. D{\bf 93}, 114014 (2016).
\bibitem{rennecke16} F. Rennecke and B.-J. Schaefer, Phys. Rev. D{\bf 96}, 016009 (2017).
\bibitem{gomeznicola16} A. Gomez Nicola and J. Ruiz de Elvira, Phys. Rev. D{\bf 98}, 014020 (2018).
\bibitem{gomeznicola21} A. Gomez Nicola, J. Ruiz de Elvira, A. Vioque-Rodriguez, and D. Alvarez-Herrero, Eur. Phys. J. C{\bf 81}, 637 (2021).
\bibitem{tiwari23} V.-K. Tiwari, Phys. Rev. D{\bf 108}, 074002 (2023).
\bibitem{rai20} S. K. Rai and V. K. Tiwari, Eur. Phys. J. Plus {\bf 135}, 844 (2020).
\bibitem{li20} X. Li, W.-J. Fu, and Y.-X. Liu,  Phys. Rev. D{\bf 101}, 054034 (2020).
\bibitem{ishii16} M. Ishii, K. Yonemura, J. Takahashi, H. Kouno, and M. Yahiro, Phys. Rev. D{\bf 93}, 016002 (2016).
\bibitem{ishii17} M. Ishii, H. Kouno, and M. Yahiro, Phys. Rev. D{\bf 95}, 114022 (2017).
\bibitem{bottaro20} S. Bottaro and E. Meggiolaro,  Phys. Rev. D{\bf 102}, 014048 (2020).
\bibitem{horvatic19} D. Horvatic, D. Kekez, and D. Klabucar, Phys. Rev. D{\bf 99}, 014007 (2019).
\bibitem{horvatic20} D. Horvatic, D. Kekez, and D. Klabucar, Eur. Phys. J. {\bf 229}, 3363 (2020).
\bibitem{gomeznicola19} A. Gomez Nicola, J. Ruiz De Elvira, and A. Vioque-Rodriguez, J. High Energy Phys. 11 (2019) 086.
\bibitem{kawaguchi23} M. Kawaguchi and D. Suenaga, J. High Energy Phys. 08 (2023) 189.
\bibitem{cuteri21} F. Cuteri, O. Philipsen, and A. Sciarra, J. High Energy Phys. 11 (2021) 141 .
\bibitem{dini22} L. Dini, P. Hegde, F. Karsch, A. Lahiri, C. Schmidt, and S. Sharma, Phys. Rev. D{\bf 105}, 034510 (2022).
\bibitem{bernhardt23} J. Bernhardt and C.-S. Fischer, Phys. Rev. D{\bf 108}, 114018 (2023).
\bibitem{chandrasekharan07} S. Chandrasekharan, A.-C. Mehta, Phys. Rev. Lett. {\bf 99}, 142004 (2007).
\bibitem{pisarski84} R. D. Pisarski and F. Wilczek, Phys. Rev. D{\bf 29}, 338 (1984).
\bibitem{fejos22} G. Fejos, Phys. Rev. D{\bf 105}, L071506 (2022).
\bibitem{fejos16} G. Fejos and A. Hosaka, Phys.  Rev.  D{\bf 94}, 036005 (2016).
\bibitem{fejos21} G. Fejos and A. Patkos, Phys. Rev. D{\bf 105}, 096007 (2022).
\bibitem{pisarski20} R. D. Pisarski and F. Rennecke,  Phys.  Rev.  D{\bf 101}, 114019 (2020).
\bibitem{grahl13} M. Grahl and D.-H. Rischke, Phys. Rev. D{\bf 88}, 056014 (2013).
\bibitem{fejos14} G. Fejos, Phys. Rev. D{\bf 90}, 096011 (2014).
\bibitem{pisarski24} R.-D. Pisarski and F. Rennecke, arXiv:2401.06130.
\bibitem{wetterich93} C. Wetterich, Phys.  Lett. B{\bf 301}, 90 (1993).
\bibitem{morris94} T. R. Morris, Int. J. Mod. Phys. A{\bf 09}, 2411 (1994).
\bibitem{litim01} D. F. Litim, Phys. Rev. D{\bf 64}, 105007 (2001).
\bibitem{mitter15} M. Mitter, J.-M. Pawlowski,  and N. Strodthoff,  Phys. Rev. D{\bf 91}, 054035 (2015).
\bibitem{shuryak94} E. Shuryak and M. Velkovsky, Phys. Rev. D{\bf 50}, 3323 (1994).
\bibitem{aoki06} Y. Aoki, Z. Fodor, S. D. Katz, and K. K. Szabo, Phys. Lett. B{\bf643}, 46 (2006). 
\bibitem{meggiolaro23} E. Meggiolaro, arXiv:2310.10339.
\bibitem{jido19} D. Jido, H. Masutani, and S. Hirenzaki,  Prog. Theor. Phys. {\bf 2019}, 053D02 (2019).
\bibitem{sakai23} S. Sakai and D. Jido, Phys. Rev. C{\bf 107}, 025207 (2023).

\end{thebibliography}
\end{document}